\documentclass[prd,twocolumn,floats,floatfix,showpacs,nofootinbib]{revtex4}
\usepackage{graphicx}
\usepackage{dcolumn}
\usepackage{bm}
\usepackage{graphics}
\usepackage{slashed}
\usepackage{amssymb}
\usepackage{natbib}
\usepackage{amsmath}
\usepackage{url}
\usepackage{color}
\newcommand{\bea}{\begin{eqnarray}}
\newcommand{\ena}{\end{eqnarray}}
\newcommand{\bean}{\begin{eqnarray*}}
\newcommand{\enan}{\end{eqnarray*}}

\begin{document}

\title{A mechanism to generate mass: the case of fermions}

\author{M. Novello\footnote{M. Novello is Cesare Lattes ICRANet Professor}} \email{novello@cbpf.br}

\affiliation{Instituto de Cosmologia Relatividade Astrofisica ICRA -
CBPF\\ Rua Dr. Xavier Sigaud, 150, CEP 22290-180, Rio de Janeiro,
Brazil}

\date{\today}

\begin{abstract}
In a recent paper \cite{novello} we have presented a mechanism to
generate mass from gravitational interaction, based on the Mach
principle, according to which the inertia of a body is a property of
matter as well as of the background provided by the
rest-of-the-universe. In \cite{novello} we realized such an idea for
a scalar field treating the rest-of-the-universe in its vacuum
state. In the present paper we describe a similar mechanism for
fermions.
\end{abstract}

\maketitle

\section{Introduction}

In the realm of high-energy physics, the Higgs model produced an
efficient scenario for generating mass to the vector bosons
\cite{halzen}.

 At its  origin appears a process relating the
transformation of a global symmetry into a local one and the
corresponding presence of vector gauge fields.

 This mechanism appeals to the intervention of a scalar field
 that appears as the vehicle which provides mass to the  gauge
 vector field  $ W_{\mu}.$  For the mass to be a real and constant value (a different value
for each field) the scalar field $ \varphi$  must be in a minimum
state of its potential. This fundamental state of the
self-interacting scalar field has an energy distribution given by $
T_{\mu\nu} = V(\varphi_{0}) \, g_{\mu\nu}.$ A particular form of
self-interaction  of the scalar field $ \varphi$ allows the
existence of a constant value $ V(\varphi_{0}) $ that is directly
related to the mass of $ W_{\mu}.$ This scalar field has its own
mass, the origin of which rests unclear. In \cite{novello} we
presented a new mechanism depending on the gravitational
interaction, that can provides mass to the scalar field. In the
present paper we extend this strategy to the case of fermions.

Although the concept of mass pervades most of all analysis involving
gravitational interaction, it is an uncomfortable situation that
still to this day there has been no successful attempt to derive a
mechanism by means of which mass appears as a direct consequence of
a dynamical process depending on gravity \cite{narlikar}.

The main idea concerning inertia in the realm of gravity according
to the origins of General Relativity, goes in the opposite
direction: while the Higgs mechanism explores the reduction of a
global symmetry into a local one, the Mach principle suggests a
cosmical dependence of local properties, making the origin of the
mass of a given body to depend on the structure of the whole
universe. In this way, there ought to exist a mechanism by means of
which this quantity - the mass -- depends on the state of the
universe. The purpose of this paper is to exhibit an example of such
mechanism.

We start by considering Mach principle as the statement according to
which the inertial properties of a body $\mathbb{A }$ are determined
by the energy-momentum throughout all space. How could we describe
such universal state that takes into account the whole contribution
of the rest-of-the-universe onto $\mathbb{A }$ ? There is no simpler
way than to consider this state as the most homogeneous one and
relate it to what Einstein attributed to the cosmological constant
or, in modern language, the vacuum of all remaining bodies. This
means to describe the energy-momentum distribution of all
complementary bodies of $\mathbb{A }$ in the Universe as
$$ T^{U}_{\mu\nu} = \Lambda
\, g_{\mu\nu}
$$

This article is written in the following stages: in the next section
I will do a revision of the mechanism in the case of matter being
represented by a scalar field. We shall see how the existence of an
action by the rest-of-the-universe on a massless field may provide
it with a mass. In the following section we will turn our attention
to the case of a massless spinor field. We end with some comments on
this strategy and further consequences.

\section{The case of scalar field}

Let $\varphi $ be a massless field the dynamics of which is given by
the Lagrangian $$ L_{\varphi} = \frac{1}{2} \,
\partial_{\alpha} \varphi \, \partial^{\alpha} \varphi$$
In the framework of General Relativity its gravitational interaction
will be given by the Lagrangian

\begin{equation}
L = \frac{1}{\kappa} \, R + \frac{1}{2} \, W(\varphi) \,
\partial_{\alpha} \varphi \, \partial^{\alpha} \varphi + B(\varphi)
\, R - \frac{1}{\kappa} \, \Lambda \label{210}
\end{equation}
where for the time being the dependence of $ B $ and $ W $ on the
scalar field is not fixed. This dynamics represents a scalar field
coupled non-minimally with gravity. The cosmological constant is
added by the reasons presented above and as we shall see it
represents the influence of the rest-of-the-universe on $\varphi.$
In the present proposed mechanism, such $ \Lambda $ is the real
responsible to provide mass for the scalar field. This means that if
we set $ \Lambda = 0$ the mass of the scalar field will vanish, as
we shall see. The factor $ W(\varphi)$ has been introduced to
eliminate terms involving  $ (\partial_{\alpha} \varphi)^{2}$ that
have their origin in the non-minimal coupling with gravity.

Independent variation of $\varphi$ and $g_{\mu\nu}$ yields
\begin{equation}
W \, \Box \varphi + \frac{1}{2} \, W' \, \partial_{\alpha} \varphi
\, \partial^{\alpha} \varphi - B' \, R  = 0 \label{211}
\end{equation}

\begin{equation}
\alpha_{0} \, ( R_{\mu\nu} - \frac{1}{2} \, R \, g_{\mu\nu} ) = -
T_{\mu\nu}
 \label{212}
\end{equation}
where for graphical simplicity we set  $ \alpha_{0} \equiv 2 /
\kappa $ and  $ B' \equiv \partial B / \partial \varphi.$ The
energy-momentum tensor defined by

 $$T_{\mu\nu} = \frac{2}{\sqrt{- g}} \, \frac{\delta ( \sqrt{-g} \,
 L)}{\delta g^{\mu\nu}} $$

is given by
\begin{eqnarray}
T_{\mu\nu} &=& W \, \partial_{\mu} \varphi \, \partial_{\nu} \varphi
- \frac{1}{2} \, W \, \partial_{\alpha} \varphi \, \partial^{\alpha}
\varphi \, g_{\mu\nu} \nonumber \\
&+& 2 B \, ( R_{\mu\nu} - \frac{1}{2} \, R \, g_{\mu\nu} ) \nonumber
\\
&+& 2 \nabla_{\mu} \nabla_{\nu} B - 2 \Box  B \, g_{\mu\nu} +
\frac{1}{\kappa} \,\Lambda \, g_{\mu\nu} \end{eqnarray}

Taking the trace of equation (\ref{212}) we obtain
\begin{equation}
( \alpha_{0} + 2 B ) \, R = - \, \partial_{\alpha} \varphi \,
\partial^{\alpha} \varphi \, ( W + 6 \, B'') - 6 B' \, \Box \varphi + 4 \,
\frac{\Lambda}{\kappa}
\end{equation}
where we use  $\Box B = B' \, \Box \varphi + B'' \,
\partial_{\alpha} \varphi \, \partial^{\alpha} \varphi.$

 Inserting this result on the equation (\ref{211} ) yields
\begin{equation}
\mathbb{M} \, \Box \varphi  \,  + \mathbb{N}  \, \partial_{\alpha}
\varphi \, \partial^{\alpha} \varphi -  \mathbb{Q} = 0 \label{215}
\end{equation}
where
$$ \mathbb{M} \equiv W + \frac{6 (B')^{2}}{\alpha_{0} + 2 B} $$

$$ \mathbb{N} \equiv \frac{1}{2} \, W' + \frac{B' \,( W + 6 B'')}{\alpha_{0} + 2 B} $$

$$ \mathbb{Q} = \frac{4 \, \Lambda \,B'}{\kappa \,(\alpha_{0} + 2 B)} $$

Thus, through the non-minimal coupling with the gravitational field
it follows that the scalar field acquires an effective
self-interaction.

At this point it is worth to select among all possible candidates of
$ B$ and $W $ a particular one that makes the factor on the gradient
of the field to disappear on the equation of motion by setting $
\mathbb{N} = 0. $ This condition will give $ W $ as a function of $
B :$
\begin{equation}
W = \frac{2q - 6 (B')^{2}}{\alpha_{0} + 2 B} \label{216}
\end{equation}
where $ q $ is a constant. Inserting this result into the equation
(\ref{215}) yields
\begin{equation}
\Box \varphi  - \frac{2 \, \Lambda}{q \, \kappa} \, B' = 0.
\end{equation}
At this point one is led to set
$$ B = - \frac{\beta}{4} \, \varphi^{2} $$
to obtain
 \begin{equation}
\Box \varphi + \mu^{2} \, \varphi = 0 \label{218}
\end{equation}
where \begin{equation}
 \mu^{2} \equiv \frac{\beta \, \Lambda}{q \, \kappa} \label{217}
\end{equation}
For the function $ W $ it follows
$$ W = \frac{2 \, \alpha_{0} - 3 \, \beta^{2} \,
\varphi^{2}}{2 \, \alpha_{0} - \beta \varphi^{2}}$$

in which we have choose $ 2 q = \alpha_{0}$ in order to obtain the
standard value for the dynamics in case $ \beta $ is zero.

Thus as a result of the above process we have succeeded to provide
to the scalar field a mass $ \mu $ that depends on the constant
$\beta $ and on the existence of $ \Lambda:$
\begin{equation}
\mu^{2} =  \beta \, \Lambda \label{219} \end{equation}

If $ \Lambda $ vanishes then the mass of the field vanishes. This is
precisely what we envisaged to obtain: the net effect of the
non-minimal coupling of gravity with the scalar field corresponds to
a specific self-interaction of the scalar field. The mass of the
field appears only if we take into account the existence of all
remaining bodies in the universe --- represented by the cosmological
constant --- in the state in which all existing matter is on the
corresponding vacuum. The values of different masses for different
fields is contemplated in the parameter $\beta.$

\subsection{Renormalization of the mass}

The effect of the rest-of-the-universe on a massive scalar field can
be analyzed through the same lines as above. Indeed, let us consider
the case in which the free field dynamics is given by

\begin{equation}
L = \frac{1}{\kappa} \, R + \frac{1}{2} \, \partial_{\alpha} \varphi
\, \partial^{\alpha} \varphi + B(\varphi) \, R - V(\varphi) -\Lambda
\label{210}
\end{equation}
where one is led to choose $ B = a + b \, \varphi -1/12 \,
\varphi^{2}$ to obtain after some algebraic steps

$$\Box \varphi + V_{eff}' = 0$$ where the effective potential $ V_{eff}  $ is
described in terms of $\Lambda,$ the pre-existing potential $ V $
and its derivatives. The net effect of the action of the
rest-of-the-universe is in this case to modify the potential $ V
$and to re-normalize the mass.

\section{The case of fermions}

Let us now turn our attention to the case of fermions. The massless
 theory for a spinor field is given by Dirac equation:
\begin{equation} i\gamma^{\mu} \partial_{\mu} \, \Psi  = 0 \label{221}
\end{equation}

In the framework of General Relativity its gravitational interaction
is driven by the Lagrangian (we are using units were $\hbar = c =
1)$

\begin{eqnarray}
L &=& \frac{i}{2} \bar{\Psi} \gamma^{\mu} \nabla_{\mu} \Psi -
\frac{i}{2} \nabla_{\mu} \bar{\Psi} \gamma^{\mu} \Psi \nonumber \\
&+& \frac{1}{\kappa} \, R +  V(\Phi) \, R - \frac{1}{\kappa}
 \, \Lambda \nonumber
\\ &+& L_{CT}
\label{3}
\end{eqnarray}
where the non-minimal coupling of the spinor field with gravity is
contained in the term $ V(\Phi)$ that depends on the scalar
$$ \Phi \equiv \bar{\Psi} \, \Psi$$
that preserves the gauge invariance of the theory under the map $
\Psi \rightarrow \exp(i \, \theta) \, \Psi.$ Note that the presence
of the factor on $ \Phi$ on the dynamics of $ \Psi$ breaks the
chiral invariance of the mass-less fermion, a condition that is
necessary for a mass to appear.

 For the time being the dependence of $ V $ on $ \Phi$ is not fixed. We have added
a counter-term $L_{CT}$ for reasons that will be clear later on.  On
the other hand, the form of the counter-term should be guessed, from
the previous analysis that we made for the scalar case, that is we
set
\begin{equation}
L_{CT} = H(\Phi) \, \partial_{\mu} \Phi \, \partial^{\mu} \Phi
\label{222}\end{equation}

This dynamics represents a massless spinor field coupled
non-minimally with gravity. The cosmological constant is added by
the reasons presented above and as we shall see it represents the
influence of the rest-of-the-universe on $\Psi.$

Independent variation of $\Psi$ and $g_{\mu\nu}$ yields
\begin{equation}
 i\gamma^{\mu} \nabla_{\mu} \, \Psi  +  \left( R \, V'  -
 H' \, \partial_{\mu} \Phi \, \partial^{\mu} \Phi  - 2 H \Box \Phi \right) \, \Psi = 0 \label{223}
\end{equation}

\begin{equation}
\alpha_{0} \, ( R_{\mu\nu} - \frac{1}{2} \, R \, g_{\mu\nu} ) = -
T_{\mu\nu}
 \label{224}
\end{equation}
where  $ V' \equiv \partial V / \partial \Phi.$ The energy-momentum
tensor is given by
\begin{eqnarray}
T_{\mu\nu} &=& \frac{i}{4} \, \bar{\Psi} \gamma_{(\mu} \nabla_{\nu)}
\Psi - \frac{i}{4} \nabla_{(\mu} \bar{\Psi} \gamma_{\nu)} \Psi
\nonumber \\
&+& 2 V ( R_{\mu\nu} - \frac{1}{2} \, R \, g_{\mu\nu} ) + 2
\nabla_{\mu} \nabla_{\nu} V - 2 \Box V g_{\mu\nu} \nonumber \\
&+& 2 H \, \partial_{\mu} \Phi \, \partial_{\nu} \Phi - H \,
\partial_{\lambda} \Phi \, \partial^{\lambda} \Phi \, g_{\mu\nu} +
\frac{\alpha_{0}}{2} \, \Lambda \, g_{\mu\nu} \label{225}
 \end{eqnarray}

Taking the trace of equation (\ref{224}) we obtain after some
algebraic manipulation:
\begin{eqnarray}
( \alpha_{0} + 2 V + V') \, R &=& H'\, \Phi  \,\partial_{\alpha}
\Phi \,
\partial^{\alpha} \Phi \nonumber \\
&+& 2 H \, \Phi \, \Box \Phi - 6 \Box V + 2 \, \alpha_{0} \, \Lambda
\end{eqnarray}

Inserting this result back on the equation (\ref{223}) yields
\begin{equation}
 i\gamma^{\mu} \nabla_{\mu} \, \Psi  +
\left( \mathbb{X} \,  \partial_{\lambda} \Phi \, \partial^{\lambda}
\Phi + \mathbb{Y} \, \Box \Phi \, \right) \, \Psi + \mathbb{Z} \,
\Psi = 0 \label{226}
\end{equation}
where
$$ \mathbb{Z} \equiv \frac{2 \, \alpha_{0} \, \Lambda \, V'}{\alpha_{0} + 2 V + \Phi \, V'}
$$

$$ \mathbb{X} = \frac{V'\, ( \Phi \, H' - 2 H - 6 V'')}{\alpha_{0} + 2  V + \Phi \, V'}
 - H' $$

$$  \mathbb{Y} = \frac{V'\, ( 2 H \, \Phi - 6 V')}{\alpha_{0} + 2  V + \Phi \, V'}
 - 2 H  $$

At this stage it is worth selecting among all possible candidates of
$ V$ and $ H $ particular ones that makes the factor on the gradient
and on $ \Box $ of the field to disappear from equation (\ref{226}).
The simplest way is to set  $ \mathbb{X} = \mathbb{Y} = 0$ which
imply only one condition, that is

\begin{equation}
H = \frac{- \, 3 (V')^{2}}{\alpha_{0} + 2 V} \label{227}
\end{equation}

The non-minimal term $ V $ is such that $ \mathbb{Z}$ reduces to a
constant, that is
\begin{equation}
V = \frac{\alpha_{0}}{2} \, \left[ (1 + \sigma \, \Phi)^{-2} - 1
\right] \label{228}
\end{equation}
Then it follows immediately
\begin{equation}
H = - 3 \alpha_{0} \, \sigma^{2} \,  (1 +  \sigma \, \Phi)^{-4}
\label{229}
\end{equation}
where $ \sigma $ is a constant.

The equation for the spinor becomes
\begin{equation} i\gamma^{\mu} \nabla_{\mu} \, \Psi  - m \Psi= 0 \label{15}
\end{equation}
where \begin{equation}
 m = \frac{4 \, \sigma \, \Lambda}{\kappa}.
 \label{20}
 \end{equation}

Thus as a result of the above process we have succeeded to provide
to the spinor field a mass $ m $ that depends crucially on the
existence of $ \Lambda.$ If $ \Lambda $ vanishes then the mass of
the field vanishes. This is precisely what we envisaged to obtain:
the net effect of the non-minimal coupling of gravity with the
spinor field corresponds to a specific self-interaction. The mass of
the field appears only if we take into account the existence of all
remaining bodies in the universe --- represented by the cosmological
constant --- in the state in which all existing matter is on the
corresponding vacuum. The values of different masses for different
fields are contemplated in the parameter $ \sigma.$

The various steps of our mechanism can be synthesized as follows:
\begin{itemize}
\item{The dynamics of a massles spinor field $ \Psi$ is described
by the Lagrangian
$$ L_{D} = \frac{i}{2} \bar{\Psi} \gamma^{\mu} \nabla_{\mu} \Psi -
\frac{i}{2} \nabla_{\mu} \bar{\Psi} \gamma^{\mu} \Psi ; $$}
\item{Gravity is described in General Relativity by the scalar of
curvature $$ L_{E} = R ;$$ }
\item{The field interacts with gravity in a non-minimal way described
by the term
$$ L_{int} = V(\Phi) \, R $$
where $ \Phi = \bar{\Psi} \, \Psi ;$ }
\item{The action of the rest-of-the-universe on the spinor field,
through the gravitational intermediary, is contained in the form of
an additional constant term on the Lagrangian noted as $ \Lambda ;$
}
\item{A counter-term depending on the invariant $ \Phi $
is introduced to kill extra terms coming from gravitational
interaction;}
\item{As a result of this process, after choosing $ V $ and $ H $ the field acquires a mass being
described  as
$$ i\gamma^{\mu} \nabla_{\mu} \, \Psi  - m \Psi= 0   $$ where
$ m $ is given by equation (\ref{20}) and is zero only if the
cosmological constant vanishes.}
\end{itemize}

 This procedure allows us to state that the mechanism proposed here is to be
understood as a form of realization of Mach principle according to
which the inertia of a body depends on the background of the
rest-of-the-universe. Besides, our strategy can be applied in a more
general context in support of the idea that (local) properties of
microphysics may depend on the (global) properties of the universe.
We will come back to this in a future paper \cite{novello2}.

\section{acknowledgements}
I would like to thank FINEP, CNPq and Faperj for financial support.
I would like also to thank Remo Ruffini for the creation of a very
pleasant scientific atmosphere at ICRANet in Pescara where this
paper was concluded.


\begin{thebibliography}{50}
\bibitem{novello2} M Novello:The cosmological origin of Heisenberg-Nambu-Jona-Lasinio non-linear
spinor interaction (to be submited)
\end{thebibliography}
\end{document}